\documentclass[aps,prd,superscriptaddress,nofootinbib,showpacs,twocolumn,reprint]{revtex4-1}
\usepackage{amsmath,amssymb}
\usepackage{epsfig,graphicx}
\usepackage{color}
\usepackage{nccmath}

\renewcommand{\thesubsection}{\arabic{section}.\arabic{subsection}}
\catcode`@=11
\@addtoreset{equation}{section}
\catcode`@=12

\begin{document}
\title{Inflation in generalized unimodular gravity}
\author{A. O. Barvinsky}
\email{barvin@td.lpi.ru}
\affiliation{Theory Department, Lebedev
Physics Institute,
Leninsky Prospect 53, Moscow 119991, Russia}

\author{N.Kolganov}
\email{nikita.kolganov@phystech.edu}
\affiliation{Moscow Institute of Physics and Technology, Institutskii per. 9, Dolgoprudny 141700, Russia}
\affiliation{Bogoliubov Laboratory of Theoretical Physics, Joint Institute for Nuclear Research, \\Joliot-Curie 6, Dubna 141980, Russia}

\begin{abstract}
The recently suggested generalized unimodular gravity theory, which was originally put forward as a model of dark energy, can serve as a model of cosmological inflation driven by the effective perfect fluid -- the dark purely gravitational sector of the theory. Its excitations are scalar gravitons which can generate, in the domain free from ghost and gradient instabilities, the red tilted primordial power spectrum of CMB perturbations matching with observations. The reconstruction of the parametric dependence of the action of the theory in the early inflationary Universe is qualitatively sketched from the cosmological data. The alternative possibilities of generating the cosmological acceleration or quantum transition to the general relativistic phase of the theory are also briefly discussed.
\end{abstract}

\maketitle

\section{Introduction}
Cosmological acceleration phenomenon and fundamental problems of quantum gravity produce a rich playground for modifications of Einstein general relativity (GR). Quite curiously, these modifications associated nowadays with the resolution of the dark energy problem were initiated by Einstein himself \cite{Einstein} in the form of what is presently known as unimodular gravity (UMG). UMG is characterized by the kinematical restriction on metric coefficients $\det g_{\mu\nu}=-1$, and despite the lack of general coordinate invariance it has attracted a lot of attention \cite{UMG,*UMG1,*UMG2}, in particular, because it incorporates the cosmological constant as an arbitrary constant of motion rather than a fixed fundamental parameter in the action. Another type of modifications is focused on the construction of UV consistent renormalizable quantum gravity and is based on breakdown of Lorentz invariance in the form of Horava-Lifshitz gravity theories \cite{Horava}. What these theories share in common with UMG is that their most advanced class, with firmly established renormalizability and possible asymptotic freedom \cite{BBHSS,*BBHSS1}, belongs to the so-called projectable models with another type of metric restriction -- fixing the lapse function $(-g^{00})^{-1/2}=1$.

Recently there was suggested another model, motivated by the necessity to explain the dark energy mechanism with a variable in time equation of state \cite{darkness}. It combines the violation of both diffeomorphism and relativistic symmetries and generalizes the kinematical restrictions of the above two types. This is generalized unimodular gravity (GUMG) in which the lapse function is identified with some rather generic function of the determinant of the 3-dimensional metric $\gamma_{ij}\equiv g_{ij}$ \cite{darkness},
    \begin{equation}
    (-g^{00})^{-1/2}=N(\gamma),
    \quad\gamma=\det \gamma_{ij}.               \label{condition}
    \end{equation}

As shown in \cite{darkness} this restriction, which obviously reduces to the unimodular condition for a particular choice of $N(\gamma)=1/\sqrt\gamma$, leads to Einstein equations with the effective matter sources -- the perfect fluid with the variable in time equation of state $p=w\varepsilon$ and barotropic parameter $w$. It turned out that this model represents a complicated example of constrained dynamical system subject to bifurcation \cite{GUMG} -- in the wording of \cite{Henneaux-Teitelboim} the existence of two branches of the theory with different numbers of degrees of freedom. One of these branches has extra degree of freedom which corresponds to the conformal mode of the spatial metric and manifests itself as effective perfect fluid. There is a large class of functions $N(\gamma)$ for which the excitations of this fluid -- scalar gravitons -- are free from ghost and gradient instabilities and propagate with a nontrivial speed of sound. This is a rather unusual property, because the scalar degree of freedom composed of the spacetime metric as a rule suffers from ghost instabilities (with, perhaps, one exception possible for  Starobinsky model of the higher-derivative $R^2$-gravity \cite{Starobinsky_model,*Starobinsky_model1}).

It turned out, however, that despite original motivation GUMG model cannot generate a cosmological acceleration scenario of crossing the phantom divide line $w=-1$ \cite{PDL}, because in the domain of stability of the theory the time dependent parameter $w$ always evolves away from $-1$ \cite{GUMG}. Since this range of $w$ below $-1$, which was once rather popular \cite{PDL,Vikman}, is still not ruled out by observations at small redshifts, GUMG theory can hardly be responsible for the dark energy mechanism. But it was conjectured in \cite{GUMG} that GUMG scalar graviton can generate phenomenologically acceptable inflationary stage. Here we indeed confirm this conjecture and show that the cosmological perturbation theory for quasi-exponential expansion of the Universe in GUMG model is very similar to the formalism of general relativistic inflationary cosmology \cite{Mukhanov_etal,*Mukhanov_etal1,Mukhanov,Garriga_etal}. Contrary to GR inflation in the GUMG model does not need any additional field like inflaton, because inflationary expansion is driven by the scalar graviton, which leads under an appropriate choice of the function $N(\gamma)$ to nearly flat power spectra of scalar and tensor perturbations with the parameters close to observations.

The organization of the paper is as follows. In Sects.\ref{section2} and \ref{section3} we briefly recapitulate the foundations of the GUMG model and its Friedmann background solutions presented in \cite{darkness,GUMG}. In particular, we derive the Friedmann equation for this background as an integral of motion of $ij$-components of modified Einstein equations like it has been done in projectable Horava-Lifshitz gravity in \cite{Mukohyama}. In Sects.\ref{section4} and \ref{section5} we develop the cosmological perturbations theory on this background and derive the expressions for power spectra of scalar and tensor perturbations. These expressions reveal a remarkable similarity with the formalism of GR inflationary cosmology driven by the inflaton scalar field in the representation of a dynamical perfect fluid with a generic equation of state \cite{Mukhanov,Garriga_etal}. Sect.\ref{section6} is devoted to the discussion of approximate reconstruction of the function $N(\gamma)$ from the known parameters of red tilted inflationary power spectra. Concluding section contains discussion and speculations on the role of GUMG theory in realistic cosmology with matter fields, undergoing transitions between various epochs which include the cosmological acceleration stage. In particular, we speculate on the possibility of classically forbidden quantum transitions between bifurcating branches of the GUMG theory that could bring it into a partially gauge fixed GR phase. Two abstracts contain the formalism of gravitational gauge invariant potentials and inclusion of matter sources into GUMG theory.

\section{Generalized unimodular gravity}\label{section2}
Dynamics of the generalized unimodular gravity \cite{darkness} is described in much detail in \cite{GUMG}. The action of this theory follows from the Einstein-Hilbert action $S_{\mathrm{EH}}[g_{\mu\nu}]$ by the substitution of the kinematical restriction (\ref{condition}). This action generates the equations of motion which effectively coincide with Einstein equations in the presence of a perfect fluid,
    \begin{align}
    &R_{\mu\nu} - \frac12\, g_{\mu\nu} R
    = \frac12\, T_{\mu\nu},              \label{Ee}\\
    &T_{\mu\nu} = (\varepsilon + p)
    \, u_\mu u_\nu + p \, g_{\mu\nu}.     \label{stress}
    \end{align}
The stress tensor of this effective perfect fluid has a 4-velocity $u_\mu =-\delta^0_\mu\,N$ and the equation of state for its energy density $\varepsilon$ and pressure $p$,
    \begin{equation}
    p = w \varepsilon, \quad w
    = 2\,\frac{d \ln N(\gamma)}{d \ln \gamma},  \label{w}
    \end{equation}
its barotropic parameter $w$ being determined by the function $N(\gamma)$.

These equations can be easily derived by treating all ten $g_{\mu\nu}$ in the variational procedure as independent, but including the kinematical constraint (\ref{condition}) in the action with the Lagrange multiplier \cite{darkness,GUMG}. It is important that the energy density $\varepsilon$ here is in fact composed of the metric, because the 00-component of the ten effective Einstein equations is just a definition of $\varepsilon$,
    \begin{equation}
    \varepsilon=2\,u^\mu u^ \nu \, G_{\mu\nu}, \quad
    G_{\mu\nu} = R_{\mu\nu} - \frac12\, g_{\mu\nu} \, R,    \label{epsilon}
    \end{equation}
as it follows from the contraction of (\ref{Ee}) with $u^\mu u^ \nu$. Therefore, the ten equations (\ref{Ee}) are not independent, but represent the projection of the vacuum Einstein equations on the set of nine independent equations as it should be for nine independent metric coefficients $g_{ij}$ and $g_{0i}$.\footnote{\label{footnote1}This, in particular, explains why every solution of vacuum Einstein theory is also a solution of the GUMG model.}

The dynamical content of the theory can be revealed by its canonical formalism which is usually described in terms of the ADM (3+1)-decomposition of the spacetime metric into the lapse function $N$ (for it we use the same notation as the function $N(\gamma)$ in (\ref{condition})), shift functions $N_i$ and the spatial metric $\gamma_{ij}$ of spacelike slices of a constant time $x^0=t$
    \begin{equation}
    \begin{aligned}
    &N = (-g^{00})^{-1/2}, \quad N_i = g_{0i},
    \quad N^i = \gamma^{ij} N_j, \\
    &\gamma_{ij}= (\gamma^{ij})^{-1} = g_{ij}.
    \end{aligned}     \label{ADMvariables}
    \end{equation}
With this parametrization the Lagrangian action of the theory reads
    \begin{align}
    &S_{\mathrm{GUMG}}[\gamma_{ij}, N^i]\nonumber\\
    &\,\,= \int dt\, d^3x \, N \sqrt{\gamma} \,
    \bigl( {}^3 \! R + K_{ij}^2
    - K^2\bigr)\Bigr|_{N=N(\gamma)},     \label{GUMG_ADM_fixed}\\
    &K_{ij} = \frac1{2N} (\nabla_i N_j + \nabla_j N_i
    - \dot \gamma_{ij}), \quad K = \gamma^{ij} K_{ij}, \label{K_def}
    \end{align}
in terms of the extrinsic curvature $K_{ij}$ and scalar curvature ${}^3 \! R$ of 3-dimensional spacelike slices of constant $t$. The Legendre transform with respect to metric coefficient velocities $\dot\gamma_{ij}$ converts this action into the canonical form
    \begin{align}
    &S[\gamma_{ij},\pi^{ij},N^i,P_i,v^i] = \int dt \, d^3x \, \bigl(\pi^{ij} \dot \gamma_{ij} + P_i \dot N^i\nonumber\\
    &\qquad\qquad\qquad\qquad- N H_\bot - N^i H_i
    - v^i P_i \bigr),                    \label{action_T}
    \end{align}
where the canonical momenta $\pi^{ij}$ and $P_i$ are conjugated respectively to $\gamma_{ij}$ and $N^i$, the Hamiltonian is a linear combination of the constraint functions of general relativity -- the Hamiltonian constraint $H_\bot$ and momentum constraints $H_i$,
    \begin{align}
    &H_\bot=\!\frac{\gamma_{im}\gamma_{jn}\pi^{ij}\pi^{mn}\! - \frac12\,\pi^2}{\sqrt\gamma}\!-\!\sqrt{\gamma} \, {}^3 \! R,\,\,
    \pi=\gamma_{ij}\pi^{ij}, \\
    &H_i= -2\gamma_{ij}\nabla_k \pi^{jk},
    \end{align}
and the primary constraints $P_i=0$ are included into the integrand of (\ref{action_T}) with the Lagrange multipliers $v^i$.\footnote{Both actions (\ref{GUMG_ADM_fixed}) and (\ref{action_T}) contain also surface terms which make their variational procedure consistent under fixed boundary conditions at spatial boundary. These terms and their role in gauge invariance properties of the theory were considered in \cite{GUMG}, but they will not be important in what follows.} Everywhere here and in what follows the lapse function coincides with the function $N(\gamma)$ of Eq.(\ref{condition}), $\nabla_i$ is the covariant derivative with respect to the metric $\gamma_{ij}$ and ${}^3 \! R$ is its scalar curvature.

According to the Dirac procedure for constrained dynamical systems, the conservation of the primary constraint $P_i=0$ leads to the sequence of three generations of constraints
    \begin{align}
    &H_i = 0,                                      \label{secondary}\\
    &\partial_i T=0,\quad T\equiv w N H_\bot,      \label{tertiary}\\
    &T\partial_iS=0.                               \label{quarternary}
    \end{align}
The secondary constraints coincide with the general relativistic (GR) momentum constraints $H_i$, the tertiary constraints reduce to the components of the gradient of a special function $T$ proportional to the GR Hamiltonian constraint and the quaternary constraint factorizes as a product of $T$ and the gradient of another function $S$ defined by
    \begin{align}
    &S =\varOmega\,\partial_k N^k
    - \frac{d \ln w}{d \ln \gamma}
    \frac{\pi N}{\sqrt{\gamma}},            \label{S}\\
    &\varOmega= 1 + w + 2 \frac{d \ln w}{d \ln \gamma}.   \label{Omega}
    \end{align}

It is worth discussing the origin of these constraints from the Lagrangian formalism. All secondary constraints and their higher generations are the consequence of equations of motion (\ref{Ee}). First of all, it is well known that with the Lagrangian values for canonical momenta $\pi^{ij}_0= - \sqrt{\gamma} (K^{ij} - \gamma^{ij} K)$ the general relativistic Hamiltonian and momentum constraints express as the following projections of the Einstein tensor
    \begin{align}
    H_i\,|_{\pi=\pi_0}&=\sqrt\gamma\, G_{\mu i}u^\mu,\\
    H_\bot|_{\pi=\pi_0}&=-2\sqrt\gamma\, G_{\mu\nu}u^\mu u^\nu,
    \end{align}
so that the secondary constraint $H_i=0$ is satisfied in view of the Einstein equation (\ref{Ee}) and the diagonal form of the perfect fluid stress tensor, $T_{\mu i}u^\mu=0$. Similarly, in view of (\ref{epsilon})
    \begin{align}
    &H_\bot=-\varepsilon\sqrt\gamma,
    \end{align}
and the spatial projection of the stress-tensor conservation law $\nabla^\mu T_{\mu i}=\partial_i (\sqrt{\gamma} \varepsilon w N)/N\sqrt{\gamma}=0$ immediately leads to the tertiary constraint (\ref{tertiary}), $\partial_iT=0$. On the other hand, the temporal projection of this conservation law $\nabla^\mu T_{\mu\nu}u^\nu=0$ can be solved with respect to $\partial_t(N\varepsilon\sqrt\gamma)$ and substituted into $\partial_t\partial_i T=-\partial_i\partial_t(Nw\varepsilon\sqrt\gamma)$ to give $0=\partial_t\partial_i T=T\partial_i S$, where
    \begin{equation}
    S = \bigl[(1 + w) \, \partial_k N^k
    + (\partial_t - N^k \partial_k) \ln w \bigr]  \label{S1}
    \end{equation}
is the the Lagrangian form of the function $S$ -- this can be directly verified by substituting into (\ref{S}) the Lagrangian expressions $\pi^{ij}_0= - \sqrt{\gamma} (K^{ij} - \gamma^{ij} K)$ for momenta. Therefore, we recover the quaternary constraint $T\partial_i S=0$ in the Lagrangian form.

The factorization of the quaternary constraint (\ref{quarternary}) implies that the theory bifurcates into two branches. The first branch corresponds to the equation $T=0$, which is thus equivalent to the general relativistic constraint $H_\perp=0$ and automatically enforces the tertiary constraint. In the second branch the function $T\sim H_\bot=-\varepsilon\sqrt\gamma$ is spatially constant but nonzero, so that the emergence of the perfect fluid with $\varepsilon\neq 0$ results in the quaternary constraint
    \begin{equation}
    \partial_i S = 0.
    \end{equation}
For nonvanishing coefficient $\varOmega$ in Eq.(\ref{S}) the sequence of constraints then terminates, because the conservation of $S$ reduces to the equation which determines the Lagrange multipliers $v^i$.

These two branches are physically very different, because they have different sets of constraints and different numbers of degrees of freedom. According to the Dirac terminology all the constraints in the first branch $(P_i,H_i, H_\bot)$ belong to the first class, and the theory is invariant with respect to three spacetime diffeomorphisms respecting the condition (\ref{condition}). Therefore, this branch can be interpreted as general relativity within a partial gauge fixation of spacetime diffeomorphisms, corresponding to this kinematical restriction on metric coefficients.

The second, physically most interesting branch, is the one in which this restriction gives rise to the effective perfect fluid with a nonvanishing density and pressure subject to barotropic equation of state of the above type \cite{darkness}. As shown in \cite{GUMG}, its full set of constraints $(P_i,H_i,\partial_iT,\partial_i S)$ incorporates both first class and second class constraints, the first class ones forming four nontrivial linear combinations of $P_i$ and $H_i$. Contrary to the first branch, the theory is invariant only with respect to two spatial diffeomorphisms which can be realized as local canonical transformations generated by the first class constraints.\footnote{A subtle mechanism of this disparity between the number of residual diffeomorphisms remaining after the partial gauge fixation (\ref{condition}) and the number of actual gauge symmetries is explained in \cite{GUMG}. Equations of motion are of course locally invariant also with respect to the third (temporal) diffeomorphism preserving (\ref{condition}), but this diffeomorphism does not leave invariant the boundary terms of the action and, thus, changes the physical state of the system. Therefore it cannot be considered as a local gauge symmetry and does not reduce the number of degrees of freedom.} Correspondingly, this branch has only two (the number of primary first class constraints among $P_i=0$) local diffeomorphism symmetries which can be realized as canonical transformations on phase space. This enlarges the physical sector of the theory from two local general relativistic degrees of freedom to three degrees of freedom. As we will see, this extra degree of freedom in this branch -- the scalar graviton -- can generate the analogue of the inflationary cosmology with phenomenologically acceptable parameters.

Important property of the bifurcating branches is that at the classical level there are no transitions between them, because a nonzero value of the function $T$ can never evolve to zero in finite time due to the evolution law
    \begin{equation}
    \dot T=TS.           \label{dotT}
    \end{equation}

\section{GUMG cosmology}\label{section3}

Cosmological applications of GUMG imply the necessity of working in various coordinate systems, especially in closed model when the homogeneity hypersurface is a three-dimensional sphere on which $\gamma= \det \gamma_{ij}$ cannot be globally regular. On the other hand, the condition (\ref{condition}) breaks both time and space diffeomorphisms, which seemingly precludes from covariant description in the transition from one coordinate system to another. There is, however, the possibility to modify the formulation of GUMG theory to preserve a kind of 3-dimensional bimetric covariance. This is achieved by introducing in the condition (\ref{condition}) the dependence on auxiliary spatial metric $\sigma_{ij}$,
    \begin{equation}
    N(\gamma)\mapsto N(\gamma/\sigma),\quad
    \sigma=\det \sigma_{ij}.              \label{comp_mod}
    \end{equation}
Then this condition becomes a scalar with respect to {\em simultaneous} coordinate transformations of two spatial metrics $\gamma_{ij}$ and $\sigma_{ij}$ and allows one to consider the model in arbitrary spatial coordinates. The auxiliary metric can be taken time independent, but generally has curvature and involves spatial coordinates, $\sigma_{ij}=\sigma_{ij}({x})$, ${x}=x^i$. Then the above equations get modified by a simple replacement of all 3-dimensional partial derivatives with a covariant derivative for the metric $\sigma_{ij}$,
    \begin{equation}
    \partial_i\mapsto\bar\nabla_i,\quad\bar\nabla_k\sigma_{ij}=0,\quad
    \bar\nabla^i=\sigma^{ij}\bar\nabla_j,
    \end{equation}
which of course preserves this auxiliary metric, but involves nonvanishing $\bar\nabla$-derivative of the dynamical metric $\gamma_{ij}$.\footnote{It should be emphasized that the {\em bimetric} covariance does not bring in the theory local gauge invariance because an auxiliary metric is not dynamical -- it plays in the action the role of external parameter which is not subject to variations in the variational principle.} In particular, since $H_\perp$ is a scalar density, the covariant derivative $\bar\nabla_i$, which should replace $\partial_i$ in the tertiary constraint (\ref{tertiary}), reads as $\bar\nabla_i T\equiv\sqrt\sigma\,\partial_i(w N H_\bot/\sqrt\sigma)$. In the quaternary constraint (\ref{quarternary}) $\partial_i=\bar\nabla_i$ since the function $S$ defined by (\ref{S}) with $\partial_kN^k$ replaced with $\bar\nabla_kN^k$ is of course a scalar of zero weight.

In homogeneous Friedmann cosmology with the metric of positive or zero spatial curvature, $k=+1$ or $k=0$ respectively,
    \begin{equation}
    ds^2 = - N^2 dt^2+a^2(t)\sigma_{ij}dx^i dx^j, \label{Friedmann_back0}
    \end{equation}
it is natural to identify the above auxiliary metric with this 3-dimensional metric $\sigma_{ij}$ of the 3-dimensional sphere of unit radius or the flat metric. In these cases we have the variables of the cosmological background in terms of the scale factor $a(t)$
    \begin{eqnarray}
    &&\gamma_{ij}=a^2(t)\sigma_{ij}(x),\quad N=N(a),
    \quad N^i=0,\quad                          \label{Friedmann_back1}\\
    &&{}^3 \! R=\frac6{a^2}\,k,\quad
    K_{ij}=-\frac{a\dot a}N\,\sigma_{ij},
    \quad H=\frac{\dot a}{Na},                       \label{Friedmann_back2}
    \end{eqnarray}
where according to (\ref{comp_mod}) $N=N(a)$ is just a function of $a$ and $H$ denotes the {\em physical} Hubble factor of the Friedmann background -- the logarithmic derivative of the scale factor with respect to the {\em cosmic} time $\tau$, $d\tau=N\,dt$ (we hope that it will not be confused with the notation for the GR Hamiltonian and momentum constraints above).\footnote{This definition of $H$ differs from the one adopted in \cite{GUMG} where this derivative was taken with respect to the coordinate time -- this explains a certain difference of our formalism from that of \cite{GUMG}.} Equations of motion for this background read
    \begin{align}
    &\frac{\delta S_{\mathrm{GUMG}}}{\delta \gamma_{ij}}
    =N\sqrt{\sigma} a^3
    \Bigl[ 2\,\frac{\dot H}N + 3(1+w)H^2\nonumber\\
    &\qquad\qquad\qquad+ (1 + 3 w)\frac{k}{a^2}\,
    \Bigr] \sigma^{ij}=0,                         \label{hom_eq}\\
    &\frac{\delta S_{\mathrm{GUMG}}}{\delta N^i}\equiv 0,
    \end{align}
where in view of homogeneity the shift component is identically satisfied. One can check that the first equation has the integral of motion with a constant $C$
    \begin{eqnarray}
    H^2+\frac{k}{a^2}=\frac{C}{3Na^3},    \label{C}
    \end{eqnarray}
which can be interpreted as the Friedmann equation with the Hubble factor, and the energy density of GUMG perfect fluid
    \begin{eqnarray}
    \varepsilon=\frac{M_P^2C}{Na^3}      \label{epsilon_const}
    \end{eqnarray}
(the right hand side of (\ref{C}) being $8\pi G\varepsilon/3=\varepsilon/3M_P^2$ in our units with the Newton constant $G=1/16\pi$ and reduced Planck mass $M_P^2=2$). A nonzero constant $C$ is what distinguishes the $T\neq 0$ branch of the model from its general relativistic branch. Note that the origin of the Friedmann equation with a nontrivial dark fluid density as an integral of motion of the $ij$-components of modified Einstein equations is the same as in Horava-Lifshitz gravity which also does not possess the Hamiltonian constraint of the variational nature \cite{Mukohyama}. However, in contrast to \cite{Mukohyama} where $N=1$ and $\varepsilon$ is always interpreted in terms of pressureless dark matter, in GUMG theory the density (\ref{epsilon_const}) depending on the function $N(a)$ corresponds to generic equation of state.

Similarly to UMG, where the cosmological constant is a constant of integration of equations of motion, here the energy density of effective perfect fluid $\varepsilon$ is introduced from initial conditions as a constant of integration. But contrary to UMG it evolves in time. From the stress tensor conservation law it identically satisfies the evolution law,
    \begin{eqnarray}
    \frac{d\varepsilon}{da}=
    -3\,(1+w)\,\frac\varepsilon{a},        \label{epsilon_equation}
    \end{eqnarray}
entirely determined by the function $w(a)$,
    \begin{eqnarray}
    w=\frac13\,\frac{d \ln N}{d \ln a},
    \end{eqnarray}
which in its turn corresponds to the chosen function $N(a)$ (cf. Eq.(\ref{w}) -- for brevity we do not change the notation for $N$ and $w$ in the transition from their argument $\gamma/\sigma=a^6$ to the argument $a$). Note that the expression (\ref{epsilon_const}) is consistent with the corollary of Eq.(\ref{dotT}) in the comoving frame of the GUMG fluid. Indeed, for $u^k\sim N^k=0$ in view of (\ref{S1}) $S=\partial_t\ln w$, whence the ratio $T/w=-N\varepsilon\sqrt\gamma$ is a time independent constant, so that  $\varepsilon\sim 1/Na^3$.

Thus, appropriate choice of functions $N(a)$ and $w(a)$ can imitate all possible stages of the cosmological evolution. With $N\sim 1/a^3$, $w\simeq -1$ one has inflationary evolution, $N\simeq {\rm const}$, $w\simeq 0$ corresponds to pressureless dust and $N\sim a$, $w\simeq 1/3$ describes the radiation dominated Universe.

\section{GUMG cosmological perturbation theory} \label{section4}
The theory of cosmological perturbations on the Friedmann background (\ref{Friedmann_back0}) was built for GUMG model in \cite{GUMG}. Despite a big difference of GUMG model from the conventional GR, where inflationary expansion is usually driven by the additional inflaton scalar field rather than by effective perfect fluid, the formalisms of cosmological perturbations in both theories turn out to be surprisingly similar. Here we briefly recapitulate the results of \cite{GUMG} in a slightly different form by using the parametrization in terms of the {\em cosmic} conformal time $\eta$, $d\eta=N dt/a$, which simplifies comparison with inflation theory applications in cosmology \cite{Mukhanov, Garriga_etal}. We also recover explicitly the Planck mass $M_P$ in the above formalism, which corresponds to rescaling of the action (\ref{GUMG_ADM_fixed}) $S_{\mathrm{GUMG}}\mapsto (M^2_P/2)S_{\mathrm{GUMG}}$ and all canonical momenta in the above expressions by the inverse factor $2/M_P^2$.

In this parametrization the Friedmann metric is $ds^2 =a^2(\eta)\,(-d\eta^2+ \sigma_{ij} dx^i dx^j)$, and the conformal time Hubble factor ${\mathcal H}=a H$ reads
    \begin{equation}
    \mathcal H= \frac{a'}{a},\quad
    \mathcal H = a \, H.
    \end{equation}
Here and everywhere below prime denotes the derivative with respect to the conformal time, $d/d\eta=(a/N)d/dt$. Correspondingly, Eqs.(\ref{hom_eq}) and (\ref{C}) can be rewritten as
    \begin{align}
    &\mathcal H^2 + k = \frac{\varepsilon a^2}{3M^2_P}  \label{FEconf}\\
    &\mathcal H' =
    - \frac16 (1 + 3w) \frac{\varepsilon a^2}{M_P^2},    \label{REconf}
    \end{align}
where the constant of integration $C$ is reparametrized in terms of the energy density of the background solution, what we will systematically do in what follows,
    \begin{align}
    \varepsilon=\frac{M_P^2 C}{Na^3}.
    \end{align}

The perturbations of independent GUMG variables $\gamma_{ij}$ and $N^i$ on the background of a classical solution can be decomposed into the scalar, transverse vector and transverse-traceless tensor components,
    \begin{align}
    &\delta\gamma_{ij}=a^2(-2 \psi\,\sigma_{ij}
    + 2 \nabla_i \nabla_j E+2\nabla_{(i} F_{j)}+t_{ij}), \label{lin_metr}\\
    &t^i{}_i = \nabla^i t_{ij}=\nabla_i F^i = 0,\\
    &\delta N^i= (\nabla^i B+V^i)N/a, \quad \nabla_i V^i = 0.
    \end{align}
Here and in what follows all spatial indices are raised and lowered by the metric $\sigma_{ij}$ and the covariant derivatives preserving this metric, $\nabla_k\sigma_{ij}=0$, are also denoted for brevity by $\nabla_k$ without a bar (as we will not need below the covariant derivatives with respect to the full metric $\gamma_{ij}$ this should not lead to a confusion).

In terms of these perturbations the scalar, vector and tensor modes decouple in the quadratic part of the full GUMG action,
$S^{(2)}_{\mathrm{GUMG}}=S_t+S_v+S_s$ on the Friedmann background. The part $S_v$ of the transverse vector modes $F^i$ and $V^i$ vanishes on the constraints of the theory. The tensor part
    \begin{align}
    S_t =\frac{M_P^2}8 \!\int\! d\eta\, d^3 x \sqrt{\sigma} a^2
    \bigl[(t'_{ij})^2
    -(\nabla_k t_{ij})^2
    -2k\,t_{ij}^2\bigr],      \label{S_t}
    \end{align}
is the action of two graviton oscillators on the non-static Friedmann background. The scalar sector of the quadratic action is \cite{GUMG}
\begin{widetext}
    \begin{align}
    S_s =\frac{M_P^2}2&\int d\eta \, d^3 x \sqrt{\sigma} a^2
    \Bigl[-6(\psi' + \mathcal H A)^2
    -4\,(\psi'+\mathcal H A)\Delta(B -  E') \nonumber\\
    &\quad\quad
    + 2k (B -  E')\Delta(B -  E')
    - 2(\psi -2A)(\Delta + 3 k)\psi
    + \frac{a^2 \varepsilon}{M_P^2}\frac{\varOmega}{w} \, A^2 \Bigr],    \label{scalar_action1}
    \end{align}
\end{widetext}
where $A$ characterizes the perturbation of the lapse function $\delta N=NA$ and reads in terms of the scalar modes as
    \begin{equation}
    A = w (\Delta E - 3 \psi),
    \end{equation}
where $\Delta=\sigma^{ij}\nabla_i\nabla_j$ is the 3-dimensional covariant Laplacian.

The tensor graviton action (\ref{S_t}) can be directly used in the studies of tensor perturbations, whereas the scalar action $S_s$ still requires reduction to the physical sector by solving relevant primary and higher order constraints -- linearized version of the constraints discussed in Sect.\ref{section2} This reduction, described in detail in \cite{GUMG}, is different for contributions of homogeneous and inhomogeneous modes in $S_s=S_s^{(0)}+S_s^{(>0)}$. The action of spatially constant variables $\psi_0$, $E_0$ and $B_0$  in (\ref{scalar_action1}),
    \begin{align}
    S_s^{(0)} & =\frac{3M_P^2}2\int d\eta \,d^3x\,\sqrt\sigma\, a^2\, \biggl\{ - 2 \bigl(\psi'_0-3 w \mathcal H \psi_0\bigr)^2 \nonumber\\ & \quad + \Bigl[3w\varOmega a^2 \varepsilon/M_P^2-2k(1 + 6 w)\Bigr] \psi_0^2 \bigg\}, \label{GUMG_quad_comp}
    \end{align}
similarly to the tensor case does not generate any constraints and incorporates one physical mode which is in fact the first order perturbation of the background solution of Eq.(\ref{hom_eq}) \cite{GUMG}. This single mechanical mode is a ghost because of the negative sign of its kinetic term and does not differ much from the scale factor in GR, except that $\psi_0$ in GUMG is dynamically independent and its nonvanishing constant of motion $C$ is freely specified by initial conditions.

In the sector of inhomogeneous modes the physical reduction consists in the transition from (\ref{scalar_action1}) to the canonical action of the variables $\psi$, $E$, $B$ and their conjugated momenta $\varPi_\psi$, $\varPi_E$ and $\varPi_B$
    \begin{align}
    \varPi_\psi &= - 2M_P^2 \sqrt{\sigma}\, a^2
    \bigl[3 (\psi' + \mathcal H A)
    + \Delta (B - E') \bigr],             \label{Pi_psi_zero}\\
    \varPi_E &= 2M_P^2 \sqrt{\sigma}\, a^2
    \bigl[\Delta (\psi' + \mathcal H A)
    - k\, \Delta (B - E') \bigr],       \label{Pi_E_zero}\\
    \varPi_B&=0,                               \label{Pi_B_zero}
    \end{align}
and the derivation of the chain of constraints which follow from the conservation of the primary constraint $\varPi_B=0$ -- the analogue of (\ref{secondary})-(\ref{quarternary}),
    \begin{eqnarray}
    &&\varPi_E = 0,                              \label{secondary1}\\
    &&\frac{a^2 \varepsilon}{2 M_P^2}\frac{\varOmega}w\, A
    + (\Delta+3k)\psi +
    \frac{\mathcal H}{2 a^2M_P^2}
    \frac{\varPi_\psi}{\sqrt\sigma} =0,                         \label{tertiary1}\\
    &&\mathcal H\frac{d \varOmega}{da}
    A+\frac{1}{6a^2 M_P^2}\frac{dw}{da}
    \frac{\varPi_\psi}{\sqrt{\sigma}}
    + w\varOmega\,\Delta B = 0,                  \label{quarternary1}
    \end{eqnarray}

The solution of these constraints with respect to $E$ and $B$ in terms of the remaining canonical variables $\psi$ and $\varPi_\psi$ then yields the canonical action
    \begin{align}
    S_s^{(>0)}[\psi, \varPi_\psi] =
    \!\int d\eta \, d^3 x \, \bigl( \varPi_\psi \psi'
    - \mbox{\boldmath$H$}[\psi, \varPi_\psi] \bigr)  \label{H_phys}
    \end{align}
with the physical Hamiltonian density
    \begin{align}
    &\mbox{\boldmath$H$}[\psi, \varPi_\psi] =
     -\frac{k\sqrt\sigma}{4 a^2 M_P^2} \frac{\varPi_\psi}{\sqrt\sigma} \,\hat{\mathcal O}^{-1} \frac{\varPi_\psi}{\sqrt\sigma} + a^2 M_P^2\sqrt\sigma\, \psi \,\hat{\mathcal O} \psi \nonumber\\
    &\qquad\quad + \frac{2M_P^4}{\varepsilon}\frac{w}{\varOmega}\sqrt\sigma\,\Bigl(\hat{\mathcal O} \psi + \frac{\mathcal H}{2a^2 M_P^2} \frac{\varPi_\psi}{\sqrt\sigma} \Bigr)^{2},                                     \label{scalar}\\
    &\hat{\mathcal O}\equiv\Delta+3k.
    \end{align}
Remarkably, this canonical formalism is in one to one correspondence with the canonical formalism of the physical sector in the theory of the minimally coupled inflaton $\varphi$ (cf. Eq.(B12) of \cite{Garriga_etal}) with a generic inflaton potential $V_I(\varphi)$ and the energy density of the homogeneous background $\varepsilon_I=\varphi^{\prime\, 2}/2a^2+V_I(\varphi)$. The identification relating the two formalisms is
    \begin{align}
    \varepsilon\,\frac{\varOmega}w\,\Leftrightarrow\,
    \varphi^{\prime\,2}/a^2=(1+w_I)\,\varepsilon_I,
    \end{align}
where $w_I$ is the parameter of the equation of state of the inflaton relating its energy density to its pressure $p_I=w_I\varepsilon_I=
\varphi^{\prime\,2}/2a^2-V_I(\varphi)$. Note that in the inflationary slow roll regime with both $w$ and $w_I$ close to $-1$ both of the above quantities are small and provide a big coefficient of the last term in the Hamiltonian density (\ref{scalar}) (and the relevant term in Eq.(B12) of \cite{Garriga_etal}).

For $k\neq 0$ this Hamiltonian density is spatially nonlocal and leads to the spatially nonlocal Lagrangian, but similarly to \cite{Garriga_etal} one can perform the canonical transformation to new variables
    \begin{align}
    q &= \frac{2M_P^2\,(-\hat{\mathcal O})^{1/2}}{\sqrt{(1+w)\,\varepsilon}} \left(\psi + \frac{\mathcal H}{2 a^2M_P^2} \, \hat{\mathcal O}^{-1}\frac{\varPi_\psi}{\sqrt{\sigma}}\right),    \label{can-trans1}\\
    p &= \frac{\sqrt{(1+w)\,\varepsilon}}{2M_P^2\,
    (-\hat{\mathcal O})^{1/2}}
    \left(-\frac{a^2 M_P^2\sqrt{\sigma}}{\mathcal H} \,
    \hat{\mathcal O} \psi + \frac{\varPi_\psi}2 \right),  \label{can-trans2}
    \end{align}
which converts the canonical action to such form that yields after the transition to the Lagrangian formalism the local action of the $q$-variable. Systematically using the background equations of motion (\ref{FEconf})-(\ref{REconf}), which express ${\mathcal H}$ and $\mathcal H'$ in terms of $\varepsilon$, and omitting the boundary terms caused by integration over $\eta$ by parts one finds
    \begin{align}
    &S_s[q] = \frac12 \int d\eta \, d^3x \, \sqrt{\sigma}\,
    \biggl\{ q'^2 + \frac{w(1+w)}{\varOmega}\, q \, \hat{\mathcal O} q \nonumber\\
    &\qquad\qquad\qquad\qquad\qquad
    +\left[\frac{z^2}2 + z\,\Bigl(\frac1z\Bigr)^{\!\prime\prime}\,
    \right] q^2 \biggr\},                             \label{q-action}\\
    &z^2 = (1 + w)\,\varepsilon a^2/M_P^2.
    \end{align}
The analogy with the formalism of the inflaton model then extends even further, because in this model the variable $z=\sqrt{(1+w_I)\varepsilon_I }a/M_P$ equals $\varphi'/M_P$ in full accordance with Eq.(B19) of \cite{Garriga_etal}.

There is however a big difference in spatial gradients part -- in contrast to the inflaton model with a unit speed of sound, the GUMG speed of sound is nontrivial
    \begin{equation}
    c_s^2=\frac{w(1+w)}\varOmega.     \label{sound}
    \end{equation}
This leads to the criterion of stability of the model against ghost and gradient instabilities of UV modes, derived in \cite{GUMG}
    \begin{equation}
    \frac{w}\varOmega>0, \quad 1+w>0.     \label{stability_domain}
    \end{equation}
The first condition follows from the positivity of the high energy kinetic term for $\psi$ in (\ref{scalar}) which is interpreted as absence of ghosts \cite{GUMG}, whereas the second condition rules out gradient instabilities.\footnote{In the $q$-parametrization of (\ref{q-action}) gradient and ghost instabilities mix up because the canonical transformation (\ref{can-trans1})-(\ref{can-trans2}) mixes kinetic and potential terms of the action. In particular, the second condition follows from the requirement of reality of these transformations, whereas the first condition is a corollary of $c_s^2>0$.}

Below we restrict ourselves with spatially flat Friedmann background which is most of all interesting from the viewpoint of inflation theory. This case of $k=0$  and $\sigma_{ij}=\delta_{ij}$ is much simpler because $\varepsilon=3M_P^2{\mathcal H}^2/a^2$ and direct transition from the canonical action (\ref{H_phys})-(\ref{scalar}) to the Lagrangian action of the original field $\psi$ gives
    \begin{equation}
    S_s[\psi] = \frac{M_P^2}2\int d \eta \,
    d^3x \, a^2 \, \frac{3\varOmega}{w}
    \Bigl(  \psi'^2 + \frac{w}{\varOmega}(1 + w)\,
    \psi\Delta\psi\Bigr),                       \label{action_conf}
    \end{equation}
after using again background equations of motion and freely integrating over $\eta$ by parts.

Therefore, the canonically normalized mode $\vartheta$,
    \begin{equation}
    \vartheta = \theta \, \psi, \qquad \theta^2
    = 3 a^2 M_P^2 \frac{\varOmega}{w},         \label{theta_psi}
    \end{equation}
has the action
    \begin{equation}
    S = \frac12 \int d\eta \, d^3x \Bigl( \vartheta'^2 + c_s^2 \, \vartheta \Delta \vartheta + \frac{\theta^{\prime\prime}\!\!}{\theta} \,  \vartheta^2 \Bigr),
    \end{equation}
and satisfies the well-known Mukhanov-Sasaki equation \cite{Mukhanov,Garriga_etal}
    \begin{equation}
    \vartheta'' - c_s^2 \, \Delta \vartheta - \frac{\theta^{\prime\prime}\!\!}{\theta}
    \, \vartheta = 0.                              \label{Mukh-Sas}
    \end{equation}

In the momentum space representation the propagating modes $\vartheta_{\mathbf k}(\eta)$ of this equation with the comoving wave vector ${\mathbf k}$, $-\Delta=|\,{\mathbf k}\,|^2\equiv k^2$, read in the short-wavelengths approximation ($c_s^2 k^2 \gg \theta''\!/\theta$) as positive and negative frequency basis functions
    \begin{align}
    \vartheta_{\mathbf k}^{(\mp)}(\eta) =
    \frac{C_\mp}{\sqrt{2c_s(\eta)}}
    \exp \Bigl[ \mp i k \int^\eta d\bar\eta
    \, c_s(\bar\eta) \Bigr].                  \label{short}
    \end{align}
The solution in the long-wavelengths approximation ($c_s^2 k^2 \ll \theta''\!/\theta$) reads as a sum of two modes,
    \begin{align}
    \vartheta_{\mathbf k}(\eta) = C_1 \,\theta(\eta) + C_2 \,
    \theta(\eta) \int^\eta_{\eta_0} \frac{d \bar\eta}{\theta^2(\bar\eta)}.        \label{long}
    \end{align}
For $\theta(\eta)$ sufficiently quickly growing with $\eta$ the integral in the second term approaches a constant value saturated by the behaviour of $\theta(\eta)$ at small values of its argument,\footnote{\label{foot1}Like for $\theta\sim\eta^\gamma$, $\gamma>1/2$, this integral behaves like $(1/\eta_0^{2\gamma-1}-1/\eta^{2\gamma-1})/{(2\gamma-1)}$, the second term forming the decaying mode which can be discarded at large $\eta$.} so that at late time $\vartheta_{\mathbf k}(\eta)$
asymptotes $C_{\mathbf k}\theta(\eta)$ with some other time independent coefficient $C_{\mathbf k}$. Therefore, long-wavelength modes freeze, $\psi_{\mathbf k}=\vartheta_{\mathbf k}/\theta\to C_{\mathbf k}$, and slowly vary only due to a decaying mode which is usually discarded at late stages of cosmological expansion.

\section{Inflationary power spectrum}\label{section5}

Suppose that the GUMG functions $N(a)$ and $w(a)$ are chosen so that they provide a sufficiently long quasi-exponential expansion, that is $w(a)$ is slightly higher than $-1$. This can be considered as the inflation stage generated by the global conformal mode -- the scale factor $a$ which, in contrast to GR, is a dynamical degree of freedom and whose perturbation -- homogeneous mode of the scalar graviton $\psi_0$ -- is related to the perturbation of the initial conditions for $a$ and $\dot a$ incorporating the constant $C$ in Eq.(\ref{C}). Let us find the power spectrum of inhomogeneous modes of this graviton, generated by its vacuum primordial fluctuations, and compare it with known inflationary spectra.

From the relation (\ref{theta_psi}), $\psi=\vartheta/\theta$, the two-point correlation function of the scalar graviton field reads as
    \begin{align}
    &\langle0| \hat \psi(\eta, \mathbf x) \, \hat \psi(\eta, \mathbf y) |0\rangle = \int \frac{dk}{k} \frac{\sin kr}{kr}  \frac{k^3}{2\pi^2 \theta^2(\eta)} \bigl|\vartheta_{\mathbf k}(\eta)\bigr|^2, \nonumber\\ & r \equiv |\mathbf x - \mathbf y|,
    \end{align}
so that the primordial power spectrum of $\psi$ equals
    \begin{equation}
    \delta_\psi^2(k,\eta) = \frac{k^3}{2\pi^2 } \frac{\bigl|\vartheta_{\mathbf k}(\eta)\bigr|^2}{\theta^2(\eta)},         \label{delta_psi}
    \end{equation}
where $\vartheta_{\mathbf k}(\eta)$ is the positive frequency basis function of the Mukhanov-Sasaki equation (\ref{Mukh-Sas}) for the canonically normalized field $\vartheta$, which corresponds to the choice of its initial vacuum $|0\rangle$. For presently observable large scales these modes were immensely blue-shifted at the beginning of the inflation stage. As is usually adopted \cite{Mukhanov}, for these early moments of time $\eta$ they are described by the adiabatic short-wavelength expression (\ref{short}), $\vartheta^{(-)}_{\mathbf k}(\eta)$, with the unit normalization $C_-=1$ relative to the Klein-Gordon inner product.

The basis function $\vartheta_{\mathbf k}(\eta)=\vartheta^{(-)}_{\mathbf k}(\eta)$ with a fixed comoving momentum evolves in time into the long-wavelength combination (\ref{long}) because $\theta''(\eta)/\theta(\eta)$ is rapidly growing in the course of cosmological expansion (for any power law behavior of $\varOmega/w\sim a^q$ its growth is defined by $(a^{2+q})''/a^{2+q} = (2+q) \bigl[3(1-w)/2 + q\bigr] H^2 a^2$), and the transition between the regimes (\ref{short}) and (\ref{long}) takes place at the moment $\eta_*$ of horizon crossing where $c_s^2 k^2\simeq \theta''/\theta$ or $c_s k=Ha$. Matching the short- and long-wavelength solutions at this moment, one finds the value of the normalizing constant $C_1=C_1(k)$ for a given comoving scale $k$,
    \begin{equation}
    C_1(k) \, \theta(\eta_*)  = \frac1{\sqrt{2c_s(\eta_*) k}},
    \quad c_s(\eta_*) k = H(\eta_*)\, a(\eta_*),
    \end{equation}
where we disregard an inessential phase factor and the contribution of the decaying mode discussed above in footnote \ref{foot1}. As a result the long-wavelength basis mode of the adiabatic vacuum of the scalar graviton becomes
    \begin{align}
    \vartheta_{\mathbf k}(\eta) &= \frac1{\theta \sqrt{2c_s k}}\bigg|_{c_s k = H a} \!\!\! \times \theta(\eta) \nonumber\\ &= \sqrt{\frac{w}{6\varOmega}} \frac{H}{(c_s k)^{3/2}M_P}\bigg|_{c_s k = H a} \!\!\!\times \theta(\eta),  \label{long1}
    \end{align}
where the time dependent function $\theta(\eta)$ is determined by the background function defined in Eq.(\ref{theta_psi}) and the rest is determined at the horizon crossing moment. Therefore, the factor $\theta(\eta)$ dutifully cancels out in the power spectrum of the scalar graviton fluctuations (\ref{delta_psi}) which becomes time independent and reads
    \begin{equation}
    \delta_\psi^2(k) = \frac1{12\pi^2}\sqrt{\frac{\varOmega}{w(1+w)^3}}\,
    \frac{H^2}{M_P^2}\bigg|_{c_s k = H a},
    \end{equation}
where all quantities are taken at the horizon crossing.

GUMG theory is not invariant under the full set of spacetime diffeomorphism and, moreover, in its physical sector local symmetries of the model are completely gauged out. However, phenomenologically observed anisotropy of the relic radiation is determined by the gravitational potentials, which are invariant under general coordinate transformations, rather than by the field $\psi$ itself. This is equally true for the GUMG theory, because here we assume a typical generally covariant coupling of matter to spacetime metric, and the laws of light propagation in a given metric are on equal footing invariant under generic diffeomorphisms both in GR and GUMG. In other words, the matter sector of GUMG theory -- the matter field action in the external curved space metric -- is invariant under these diffeomorphisms irrespective of the dynamics of the metric itself even despite the breakdown of coordinate invariance by the kinematical restriction (\ref{condition}). This means that for the sake of calculating observable CMBR we have to know the spectrum of invariant gravitational potentials rather than the spectrum of $\psi$-perturbations.

There are two such Bardeen invariants built entirely from metric perturbations \cite{Bardeen}
    \begin{align}
    &\varPsi = \psi - \mathcal H (B - E'), \label{inv_1}\\
    &\varPhi = A + \frac1a \frac{d}{d\eta}
    \Bigl[ a (B - E')\Bigr]. \label{inv_2}
    \end{align}
As is known, in GR on solutions of linearized equations for metric perturbations on spatially flat Friedmann background these two invariants coincide when the spatial part of the matter stress tensor is diagonal \cite{Mukhanov}. In Appendix \ref{AppA} we prove this also for GUMG theory on the vacuum $k=0$ Friedmann background and show that they express in terms of the scalar graviton function $\psi$,
    \begin{equation}
    \varPhi=\varPsi = \frac32 \frac{\varOmega\mathcal H}{w}\,
    \frac1{\Delta}\psi'.                      \label{Psi_Phi}
    \end{equation}
Moreover, along the lines of \cite{Mukhanov} we show that their long-wavelength behavior at late times, when the contribution of a decaying mode can be discarded, has even simpler representation relating the perturbations of $\varPhi$ and $\psi$ by a special constant in space and slowly varying in time factor
    \begin{equation}
    \varPhi = \psi \frac1a \frac{d}{d\eta} \biggl[\frac1{a}
    \!\int\! d\eta\, a^2 \biggr] = \psi \frac{d}{d\tau}\biggl[\frac1a \!\int\! d\tau \, a\biggr].
    \label{Phi_psi1}
    \end{equation}
This factor is an order of magnitude one quantity which for various post-inflationary stages with a power law dependence on the proper cosmic time $\tau=\int d\eta\,a$, $a(\tau)\propto\tau^p$, approximately equals $1/(p+1)$, $\varPhi = \psi/(p+1)$. Therefore, the observable long-wavelength part of the power spectrum reads
    \begin{equation}
    \delta_\varPhi^2(k, \eta) = \frac1{12\pi^2(p+1)^2}\sqrt{\frac{\varOmega}{w(1+w)^3}}\,
    \frac{H^2}{M_P^2}\bigg|_{c_s k = H a}.
    \end{equation}
In terms of the speed of sound of the scalar graviton (\ref{sound}) and the energy density $\varepsilon$ it exactly coincides with the classical result in the hydrodynamical formalism of inflation \cite{Mukhanov} including k-inflation \cite{k-inflation,*k-inflation1},
    \begin{equation}
    \delta_\varPhi^2(k, \eta) = \frac1{36\pi^2(p+1)^2}\frac1{c_s(1+w)}
    \frac\varepsilon{M_P^4}\bigg|_{c_s k = H a}.
    \end{equation}
 The corresponding spectral index, after replacing the differentiation with respect to $k=Ha/c_s$ by that of $a$,
    \begin{equation}
    n_s - 1 = \frac{d \ln \delta_\varPhi^2(k)}{d \ln k}
    =\frac1{\frac{d}{d\ln a}\bigl[\ln\frac{Ha}{c_s}\bigr]}\frac{d\ln \delta_\varPhi^2}{d\ln a}\,\bigg|_{c_sk=Ha},
    \end{equation}
reads
    \begin{equation}
    n_s - 1 =\frac{- 6 (1 + w)+\frac{d \ln \varOmega}{d \ln a}
    - \frac{d \ln w}{d \ln a}
    - 3 \frac{d \ln(1 + w)}{d \ln a}}{-(1+3w)
    + \frac{d \ln \varOmega}{d \ln a}
    -  \frac{d \ln w}{d \ln a}
    - \frac{d \ln(1 + w)}{d \ln a}}\Bigg|_{c_sk=Ha}.                  \label{n_s}
    \end{equation}
Note that the first terms in the numerator and denominator correspond to the standard result for a slow roll scalar inflaton, $n_s-1=6(1+w)/(1+3w)\simeq -3(1+w)$, while the other terms are generated by the nontrivial time dependence of the sound speed.

Analogous calculations for the tensor graviton action (\ref{S_t})
    \begin{equation}
    S_t = \frac12 \sum_{I=\pm}\int d \eta \, d^3x \, \Bigl( v_I'^2 + v_I \, \Delta v_I + \frac{a''\!\!}{a} \, v_I^2 \Bigr),
    \end{equation}
rewritten in terms of two canonically normalized field polarizations $v_I$,
    \begin{equation}
    t_{ij} = \frac2{M_Pa}\sum_{I=\pm}e^I_{ij} \, v_I, \quad e^I_{ij} \, e^{J \, ij} = \delta^{IJ},
    \end{equation}
give a standard result for the tensor primordial spectrum, because the speed of sound of tensor gravitons in GUMG model coincides with the speed of light, $c_s=1$, and the basis function of both graviton polarizations, $v_{\mathbf k}$, satisfies a typical equation
    \begin{equation}
    v_{\mathbf k}'' + \Bigl( k^2 - \mfrac{a''\!\!}{a} \,
    \Bigr) v_{\mathbf k} = 0.
    \end{equation}
Its long-wavelength modes  $v_{\mathbf k} (\eta) = a(\eta)/(\sqrt{2k}\,a_{k = Ha})$, which match with the short-wavelength modes $\exp(\mp ik\eta)/\sqrt{2k}$ at the horizon crossing $k=aH$, give the total primordial power spectrum of both polarizations
    \begin{align}
    &\langle 0 | \hat t_{ij}(\eta, \mathbf x) \, \hat t^{ij}(\eta, \mathbf y) | 0 \rangle = \int \frac{dk}{k} \frac{\sin{kr}}{kr} \, \frac{2k^3 \, |v_{\mathbf k}(\eta)|^2}{\pi^2 a^2(\eta)M_P^2},\\
    &\delta^2_t (k, \eta) = \frac2{\pi^2 a^2(\eta)M_P^2}k^3 \, |v_{\mathbf k}(\eta)|^2 = \frac2{\pi^2}\frac{H^2}{M_P^2} \bigg|_{k = Ha},
    \end{align}
and the spectral index which for small $1+w$ equals
    \begin{equation}
    n_t = \frac{d \ln \delta^2_t }{d \ln k} =  \frac{6(1+w)}{1+3w}\biggr|_{k = Ha} \!\! \simeq - 3(1 + w)\big|_{k = Ha}. 
    \end{equation}
The tensor to scalar ratio in GUMG theory depends on the ratio of energy densities at different horizon crossings for the tensor and scalar gravitons. At the radiation dominated epoch with $p=1/2$ it equals
    \begin{equation}
    r\equiv\frac{\delta_t^2}{\delta_\varPhi^2}
    =54\,\frac{H^2_{k=Ha}}{H^2_{c_s k=Ha}}\,
    \left[c_s(1+w)\bigr|_{c_s k=Ha}\right],     \label{r}
    \end{equation}
which coincides with the known expression in the inflaton driven model with $c_s=1$ \cite{Mukhanov}.

\section{Reconstruction of the model}\label{section6}
We will not try to reconstruct the GUMG functions $N(\gamma)$ and $w(\gamma)$ that would match with the observable cosmological data throughout the whole evolution of the Universe. But a remarkable similarity of the cosmological perturbation formalism in the GUMG model and GR inflationary cosmology suggests to consider a possible choice of these functions that could provide inflationary scenario with basic phenomenologically acceptable features -- a nearly flat spectrum of long-wavelength CMBR with the parameters close to observations. This might be nontrivial because of stringent restrictions imposed on these functions by the requirements of stability of the theory, on the one hand, and the necessity to match their choice with the phenomenology of the early Universe, on the other hand. The difficulty is related to a rather peculiar expression for the speed of sound of the scalar graviton (\ref{sound}) and a rather involved form (\ref{Omega}) of the function $\varOmega$ in this expression. Interestingly, despite the fact that the GUMG model is very similar to the general relativistic inflationary cosmology driven by a perfect fluid matter, the speed of sound excitations in the GUMG fluid coincides with the known hydrodynamical expression, $c_s^2=\partial p/\partial\varepsilon=d(w\varepsilon)/da/(d\varepsilon/da)+O\bigl((dw/da)^2\bigr)$, only up to the first order in the rate of change of the barotropic parameter \cite{GUMG}. Apparently, this discrepancy originates from the fact that in GUMG theory its perfect fluid is not an additional constituent of the system, but is a part of the gravitational field itself, and this makes a comparison of the GUMG model with the cosmological phenomenology rather incomplete.\footnote{In the relativistically invariant hydrodynamical version of inflation formalism \cite{Mukhanov} the speed of sound is actually determined by the dependence of $p$ and $\varepsilon$ on the relativistic invariant $X=g^{\mu\nu}\partial_\mu\phi\partial_\nu\phi$ which determines the fluid 4-velocity $\sim\partial_\mu\phi$ from the velocity potential $\phi$, $c_s^2=\partial_X p/\partial_X\varepsilon$. Lorentz noninvariant GUMG theory does not have such entities, and the role of $X$ is played by $a$ or $\gamma=a^3$, so that the analogy with hydrodynamical theory is rather incomplete. We thank A.Vikman for this observation.}

The choice of the function $N(\gamma)$ compatible with the inflation stage in the early Universe with small $\gamma=a^3$ implies the asymptotics $N(\gamma) \to 1/\sqrt{\gamma}$ at $\gamma \to 0$, corresponding to $w\to -1$ and the initial Hubble factor $H_0$ in the Friedmann equation (\ref{C}) with $C=3H_0^2$. Corrections to this law providing the growth of $w$ from $-1$ and a subsequent exit from inflation can be modelled as
    \begin{equation}
    N(\gamma) = \frac1{\sqrt{\gamma}} \Bigl[1 + A \, \Bigl(\mfrac{\gamma}{\gamma_0}\Bigr)^{\!\alpha}
    \Bigr],                                          \label{trial1}
    \end{equation}
where $A>0$ and $\alpha>0$ are some dimensionless parameters and $\gamma_0=a^3_0$ corresponds to the cosmological size at the end of inflation. With this choice of the function $N(\gamma)$ the main quantities of the theory read for $\gamma\ll\gamma_0$ as $w\simeq -1$, $c_s^2\simeq 1/(2\alpha-1)$ and $n_s-1\simeq - 6\alpha$. An obvious difficulty with this choice is that positivity of $c_s^2$ requires $\alpha > 1/2$ which implies $n_s-1<-3 + O\bigl((\gamma/\gamma_0)^\alpha \bigr)$ and therefore contradicts  a small value $n_s-1\simeq -0.04$ imposed by the known CMB data. There is, however, the possibility to select the value $\alpha = 1/2$ and include additional power term in the expression (\ref{trial1}),
    \begin{equation}
    N(\gamma) = \frac1{\sqrt{\gamma}} \Bigl[1 + A \sqrt{\mfrac{\gamma}{\gamma_0}}
    + B \, \Bigl(\mfrac{\gamma}{\gamma_0}\Bigr)^{\!\beta} \,
    \Bigr], \quad \beta > \frac12.                    \label{Nbeta}
    \end{equation}
As a result one has up to terms of higher order in powers of the ratio $\gamma/\gamma_0$
    \begin{align}
    &w\simeq -1 + A \sqrt{\mfrac{\gamma}{\gamma_0}}, \label{w_exp}\\
    &\varOmega\simeq -2 \beta (2\beta - 1) B\, \Bigl(\mfrac{\gamma}{\gamma_0}\Bigr)^{\!\beta}, \\
    &c_s^2\simeq\frac{A}{2B} \,\frac1{\beta(2\beta-1)}\, \Bigl(\mfrac{\gamma}{\gamma_0}
    \Bigr)^{\!\frac12-\beta},                       \label{csbeta}\\
    &n_s-1\simeq 3\, \frac{2\beta-3}{6\beta-1}.         \label{nsbeta}
    \end{align}
Thus, $n_s-1$ is negative for $1/2<\beta<3/2$ and small when $\beta$ tends to $3/2$. To demonstrate the difficulties in the reconstruction of $N(\gamma)$ let us first consider the case of $\beta=3/2$, when one has to take into account higher-order corrections in the expression for $N(\gamma$), in order to obtain nonvanishing spectral tilt $n_s-1$. The inclusion of extra terms in
    \begin{align}
    N(\gamma) = \frac1{\sqrt{\gamma}} \Bigl[&1 + A  \sqrt{\mfrac{\gamma}{\gamma_0}} + B \, \Bigl(\mfrac{\gamma}{\gamma_0}\Bigr)^{3/2} \nonumber\\
    &+ B_1 \, \Bigl(\mfrac{\gamma}{\gamma_0}\Bigr)^2 + B_2 \, \Bigl(\mfrac{\gamma}{\gamma_0}\Bigr)^{5/2} + \ldots \Bigr] \label{n32}
    \end{align}
produces the expressions
    \begin{align}
    w &= -1 + A \, \sqrt{\mfrac{\gamma}{\gamma_0}} + O(\gamma/\gamma_0), \\
    \varOmega &= -6 B \, \Bigl(\mfrac{\gamma}{\gamma_0}\Bigr)^{\!3/2} + O\bigl((\gamma/\gamma_0)^2\bigr), \\
    c_s^2 &= \frac{A}{6 B}
    \Bigl(\mfrac{\gamma}{\gamma_0}\Bigr)^{\!-1} + O\bigl((\gamma/\gamma_0)^{-1/2}\bigr).       \label{cs_expansion}
    \end{align}
They give rise to the parameters of the inflationary spectra which in the leading order in $\gamma/\gamma_0$ involve three subleading orders of the expansion (\ref{n32}) with the coefficients $A$, $B$ and $B_1$,
    \begin{align}
    & \delta_\varPhi^2(k) = \frac{\sqrt{6 B}}{27\pi^2 A^{3/2}} \frac{H_0^2}{M_P^2}\nonumber\\
    &\qquad\times\left[1 + \bigl(A + \mfrac{B_1}{B}\bigr) \sqrt{\mfrac{\gamma}{\gamma_0}} + O\Bigl(\mfrac\gamma{\gamma_0}\Bigr) \right] \bigg|_{c_s k = H a}, \\
    &n_s-1=\frac34\bigl(A + \mfrac{B_1}{B}\bigr) \sqrt{\mfrac{\gamma}{\gamma_0}}\,
    \Big|_{c_s k = H a}
    + O\Bigl(\mfrac\gamma{\gamma_0}\Bigr),       \label{ns4}\\
    &r = \frac{\delta_t^2(k)}{\delta_\varPhi^2(k)} \simeq \frac{54 A^{3/2}}{\sqrt{6 B}},
    \end{align}
and the condition of the horizon crossing $c_s k = H a$ becomes the equation on the scale factor \begin{equation}
\sqrt{\frac{A}{6B}} \frac{k}{H_0 a_0} \simeq \Bigl(\frac{a}{a_0}\Bigr)^4. \label{ka_rel}
\end{equation}

The full number of parameters $H_0,A,B,B_1$ is in principle sufficient to fit basic estimates coming from observations, $n_s-1\simeq -0.04$, $\delta_\varPhi^2\simeq 10^{-10}$ and $r\ll 1$ \cite{Planck,*Planck1}. However, if we also want to preserve the typically assumed duration of inflation from the moment of horizon crossing by a pivotal wavelength mode in the COBE part of CMBR spectrum,
    \begin{equation}
    \left.\frac{\gamma}{\gamma_0}\,\right|_{c_s k = H a}=e^{-6\mathcal N}, \label{e_fold}
    \end{equation}
with the e-folding number $\mathcal N\simeq 60$, then these parameters should satisfy a number of extra restrictions for they will otherwise come into contradiction with the expansion (\ref{n32}). Validity of this expansion till the exit from inflation when $\gamma\simeq\gamma_0$ and $w\simeq 0$ implies that $A,B,B_1=O(1)$, whereas $B_1/B\sim e^{3\mathcal N}$ to provide the magnitude of $n_s$. Therefore, $B$ should be exponentially small, $B\sim e^{-3\mathcal N}$. This leads to an exceedingly low amplitude $\delta^2_\varPhi\sim e^{-3\mathcal N/2}H_0^2/M_P^2$, unless $H_0$ takes a gigantic scale $H_0\sim e^{3\mathcal N/4}M_P$, and it also leads to inadmissibly high tensor to scalar ratio $r\sim e^{3\mathcal N/3}$.

To avoid these controversial estimates one can consider essentially nonanalytic function (\ref{Nbeta}) with the parameter $\beta$ slightly lower than its upper limit 3/2, $\beta=3/2-\Delta\beta$ with a small positive $\Delta\beta\simeq 0.05$.
    \begin{align}
    &\delta_\varPhi^2(k, \eta)\simeq\frac{\sqrt{6 B}}{27\pi^2 A^{3/2}} \frac{H_0^2}{M_P^2}
    \Bigl(\mfrac\gamma{\gamma_0}\Bigr)^{\!-\frac{\Delta\beta}2} \Big|_{c_s k = H a},\\
    &n_s\simeq1-3\Delta\beta/4,\\
    &r\simeq \frac{A^{3/2}}{\sqrt{6 B}} \Bigl(\mfrac\gamma{\gamma_0}\Bigr)^{\!\frac{\Delta\beta}2} \Big|_{c_s k = H a}.
    \end{align}
This would generate according to (\ref{ka_rel}) the power-law dependence of the power spectrum amplitude $\delta_\varPhi^2 \propto k^{-3\Delta\beta/4}$, which obviously leads to the needed value of the spectral index $n_s\simeq 0.96$ (cf. Eq.(\ref{nsbeta}) for general $\beta$). At the same time, usually assumed number of e-folding after the horizon crossing (\ref{e_fold}) gives sufficiently small values of the power spectrum amplitude and the tensor to scalar ratio.
Here we took into account in the calculation of $r$ by Eq.(\ref{r}) that during inflation stage $H^2_{c_sk=Ha}\simeq H^2_{k=Ha}\simeq H^2_0$, even though the horizon crossing by a tensor mode happens much earlier than by a scalar one -- the size of the Universe then is $c_s$ times smaller than that of the scalar mode horizon crossing.

In view of smallness of $\Delta\beta/2$ the factor $(\gamma/\gamma_0)^{-\Delta\beta/2}|_{c_sk=Ha}=e^{3\mathcal N \Delta\beta}\simeq 10^3$, so that the parameters $A$ and $B$ can be taken $O(1)$ to fit the observational data under a usual assumption that the inflation Hubble factor $H_0$ is several orders of magnitude below $M_P$. So we conclude that $r\sim 10^{-3}$ as it should be according to the present bounds on the amplitude of the tensor signal \cite{Planck}.

\section{Conclusions}\label{section7}
All this shows that inflation and more generally cosmological acceleration stage of the Universe can, in principle, be driven by the scalar graviton of the vacuum GUMG theory without any extra matter constituents like inflaton field. This enlarges the list of the models with a similar property, starting with UMG and including the theory of vacuum energy sequestering \cite{Kaloper_etal}, QCD holonomy mechanism of inflation and dark energy \cite{Zhitnitsky,*Zhitnitsky1} and others. The formalism for inflationary spectra is remarkably similar to the hydrodynamical version of inflation and k-inflation \cite{Mukhanov,k-inflation,*k-inflation1} and under a proper reconstruction of the function $N(\gamma)$ these spectra can match observations.

We will not try to discuss here the exit from inflation, reheating and the following radiation and matter dominated stages which are predominantly determined by matter particles created and thermalized in the end of inflation. Inclusion of matter into GUMG theory can easily be done without conceptually changing its dynamics. In GUMG cosmology with matter sources the Friedman equation arises simply by adding the matter energy density $\varepsilon_m$ to the GUMG fluid energy density $\varepsilon=CM^2_P/Na^3$ on the right hand side of (\ref{Ee}),
    \begin{align}
    &H^2+\frac{k}{a^2}=
    \frac{1}{3M_P^2}\,(\varepsilon+\varepsilon_m),   \label{Eem}
    \end{align}
and this equation, similarly to (\ref{Ee}), is the integral of motion of the $ij$-components of Einstein equations -- this is shown for an arbitrary matter field with a diagonal stress tensor in Appendix \ref{AppB}. The effect of $\varepsilon_m$ modifies the above formalism of the inflationary stage and nontrivially superimposes with the GUMG dynamics at later cosmological stages when the contribution of matter becomes strongly significant.

It should be emphasized that in pure GUMG theory the succession of radiation, matter and dark energy dominated stages would be determined entirely by the choice of the function $N(\gamma)$, whereas in conventional cosmology these stages are induced from different equations of state successively replacing one another in the course of matter phase transitions. As discussed in \cite{GUMG}, GUMG theory fails to describe the dark energy scenario if one insists on the requirement of crossing the phantom divide $w=-1$ without breaking the stability of the theory by ghost and tachyon modes. But if we retract this requirement (which anyway is not really supported at a significantly high confidence level \cite{Planck,*Planck1}) then the cosmological acceleration stage can be mimicked by GUMG theory. For this purpose $N(\gamma)$-function at large values of the scale factor $a$ should be approximated by the dominant term of (\ref{trial1}) including power corrections in $\gamma/\gamma_0\gg1$ with a  negative $\alpha$. At this epoch the influence of ordinary ever dissolving matter becomes smaller and smaller, so that the vacuum GUMG model might be behind the present day dark energy phenomenon just like it solely underlies the early inflation stage along the lines of the above type. What can be the interpolation between these two asymptotics of $N(\gamma)$ at very early and late times and how significant can be the contribution of the GUMG energy $\varepsilon=M_P^2 C/N\sqrt\gamma$ at intermediate cosmological epochs are, however, open issues which we do not address here. Still the possibility of realizing the dark energy scenario by GUMG model should not be completely ruled out.

There is, of course, another layer of questions that should be addressed within this model. They include celestial mechanics, solar system and table-top (fifth force search) tests of gravity theory which should confirm or disprove the validity of the GUMG model. Its possible justification can rely on a simple observation that the GUMG equations of motion are just the projections of the conventional Einstein equations, so that any classical solution of general relativity is simultaneously a solution of generalized unimodular gravity (cf. footnote \ref{footnote1}). However, there are additional solutions and the additional degree of freedom in GUMG theory, and it is exactly their contribution that makes it possible to drive inflation by a scalar graviton without any other matter constituents. Nontrivial effects of this degree of freedom can and should show up in present day tests at intermediate and small distance scales and can start contradicting observations.

A possible justification of the model could then come from the existence of its two bifurcating branches discussed in Sect.\ref{section2} Point is that the choice of the function $N(\gamma)$ and the corresponding $w(\gamma)$ can hardly provide stability of the dark sector of the theory throughout the whole cosmological evolution. If we abandon the idea of simulating dark energy by the GUMG dark fluid and if we assume that the succession of inflationary, radiation, matter and dark energy stages is predominantly generated by matter constituents of the Universe, the function $\varOmega=1+w+d\ln w/3d\ln a$ can still pass through zero in the course of these stages, which will result in ghost instabilities and, moreover, in the succession of strong coupling phases which would invalidate perturbation theory. As noted in \cite{GUMG}, for $\varOmega=0$ the theory possesses additional canonical constraints which eliminate the scalar graviton from the spectrum of the theory, and this elimination is mediated by the vanishing coefficient of the kinetic term of the scalar graviton, which puts the theory into nonperturbative strong coupling regime.\footnote{We thank S.Sibiryakov for the discussion of this point.} This might lead to quantum transitions between the two bifurcating branches of the GUMG theory discussed in Sect.\ref{section2}, which are classically forbidden by Eq.(\ref{dotT}). But the branch with $C=0$ is just general relativity in the temporal gauge (\ref{condition}). This opens a possible scenario alternative to the formation of cosmological acceleration stage by the GUMG dark fluid. This is a hypothetical scenario of the very early inflationary Universe starting in the GUMG phase and then jumping by a nonperturbative mechanism to the phase of Einstein theory in a special gauge (\ref{condition}) with conventional general relativistic laws. There is a lot more prospective issues to be solved within this model.

\section*{Acknowledgments} We want to express special thanks to Alexander Kamenshchik with whom the idea of the generalized unimodular gravity model was originally put forward. Also we are indebted to Sergey Sibiryakov and Alex Vikman for stimulating discussions. This work was supported by the RFBR grant No.17-02-00651 and by the Foundation for Theoretical Physics Development “Basis”.

\section*{Appendices}
\addcontentsline{toc}{section}{Appendices}
\renewcommand{\thesubsection}{\Alph{subsection}}
\numberwithin{equation}{subsection}

\subsection{Gauge invariant gravitational potentials\label{AppA}}
Here we derive onshell expressions for the Bardeen gravitational invariants \cite{Bardeen} (\ref{inv_1})-(\ref{inv_2}) in terms of physical variable $\psi$. In view of the Lagrangian expressions for momenta (\ref{Pi_psi_zero})-(\ref{Pi_E_zero}) with $k=0$ and the secondary constraint $\varPi_E=0$ one has
    \begin{align}
    &A=-\frac{\psi'}{\mathcal H} , \label{constr_cons}\\
    &\Delta(B - E') = -\frac{\varPi_\psi}{2a^2M_P^2}.
    \end{align}
On the other hand, for the action (\ref{H_phys}) with the Hamiltonian density (\ref{scalar}) the Lagrangian expression for the momentum $\varPi_\psi$ reads as
    \begin{equation}
    \frac{\varPi_\psi}{2a^2M_P^2}=-\frac{\Delta\psi}{\mathcal H}+\frac{3\varOmega}{2w}\psi'.
    \end{equation}
Using it\footnote{Note that this expression would be different for the action (\ref{action_conf}), because this action (\ref{action_conf}) differs from (\ref{H_phys}) by a surface term, and its momentum conjugated to $\psi$ is different from $\varPi_\psi$.} one obtains the onshell expression for $B-E'$ in terms of $\psi$ and $\psi'$,
    \begin{equation}
    B - E' = \frac{\psi}{\mathcal H}-\frac32 \frac{\varOmega}{w}\frac1\Delta\psi',      \label{B-A'}
    \end{equation}
and the invariant (\ref{inv_1}) takes the form
    \begin{equation}
    \varPsi = \frac32 \frac{\varOmega\mathcal H}{w}\frac1{\Delta}\psi'.
    \end{equation}

Substitution of (\ref{B-A'}) into the expression for $\varPhi$ and the use of background equations of motion to express $\mathcal H'$ in terms of $\mathcal H^2$ gives
    \begin{equation}
    \varPhi = -\frac32\frac{\varOmega\mathcal H}{w}\frac1{\Delta}\psi'
    - \frac32\frac1\Delta \biggl[\Bigl(\frac{\varOmega}{w}\psi'\Bigr)^{\!\prime} \!
    -\frac32 (1+w)\,\Delta \psi\biggr],
    \end{equation}
whence in view of the equation of motion following from the action (\ref{action_conf}) for $\psi$
    \begin{equation}
    \frac1{a^2}\Bigl(a^2\frac{\varOmega}{w}\psi'\Bigr)'
    -(1+w)\,\Delta\psi=0,                 \label{psi_equation}
    \end{equation}
one finds that both gauge invariants coincide and equal
    \begin{equation}
    \varPhi = \varPsi = \frac32 \frac{\varOmega\mathcal H}{w} \frac1{\Delta}\psi'.
    \end{equation}

After horizon crossing the long-wavelength modes of these gauge invariants can be represented in terms of $\psi$ somewhat differently \cite{Mukhanov}. Eq.(\ref{psi_equation}) can be rewritten as the following equation for the gravitational potential $\varPhi$
    \begin{equation}
    \Bigl(\frac{a}{H} \varPhi\Bigr)' = \frac32 a^2 \, (1+w) \, \psi,
    \end{equation}
whence
    \begin{equation}
    \varPhi(\eta) = \frac32 \frac{\mathcal H}{a^2} \int_{\eta_0}^\eta d \bar\eta\,a^2(\bar\eta) \, \bigl(1 + w(\bar\eta)\bigr) \, \psi(\bar\eta)
    \end{equation}
up to the contribution of the decaying mode which is proportional to $\mathcal H/a^2$ and which we will discard. As mentioned in the end of Sect.\ref{section4}, the long-wavelength perturbation $\psi$ is also frozen up to the contribution of the same mode and, therefore, it can be pulled out of the integral, so that
    \begin{equation}
    \varPhi = \psi\,\frac{\mathcal H}{a^2}\int d\eta\,a^2 \left[\Bigl(\frac1{\mathcal H}\Bigr)'+1\right],    \label{psi_1}
    \end{equation}
where we used the corollary of the background equations of motion (\ref{FEconf})-(\ref{REconf}) with $k=0$
    \begin{equation}
    1+w=\frac23\left[\Bigl(\frac1{\mathcal H}\Bigr)'+1\right]
    \end{equation}
and did not specify obvious limits of integration because of systematically disregarding the decaying mode contributions. Integrating in (\ref{psi_1}) by parts we obtain with the same precision Eq.(\ref{Phi_psi1}).

\subsection{Einstein equations in GUMG theory with matter} \label{AppB}
In the presence of matter with the diagonal stress tensor
    \begin{equation}
    T^m_{\mu\nu} = (\varepsilon_m + p_m) \, u_\mu u_\nu + p_m \, g_{\mu\nu},     \label{stressm}
     \end{equation}
which has the matter energy density $\varepsilon_m$ and pressure $p_m$, the Einstein equations of the GUMG theory have a standard form with the stress tensor (\ref{stress}) of the GUMG fluid
     \begin{equation}
     G_{\mu\nu}=\frac1{M_P^2}(T_{\mu\nu}+T^m_{\mu\nu}).
     \end{equation}
Taking their $\bot\bot$-projection and expressing the GUMG fluid density in terms of the Friedmann metric,
     \begin{equation}
     \varepsilon=M_P^2G_{\bot\bot}-\varepsilon_m,\quad
     G_{\bot\bot}\equiv 3\,\Bigl(H^2+\frac{k}{a^2}\Bigr),    
     \end{equation}
one has the $ij$-components of the equations for Friedmann metric,
    \begin{equation}
    G_{ij}=\frac1{M_P^2}\bigl(M_P^2\,w\,G_{\bot\bot}
    +p_m-w\varepsilon_m\bigr)\gamma_{ij}, \label{Eeij}
    \end{equation}
    \begin{equation}
    G_{ij}\equiv-\Bigl(\frac{2\dot H}N+3H^2+\frac{k}{a^2}\Bigr)\gamma_{ij}.
    \end{equation}    
Then the time derivative of the following linear combination of $G_{\bot\bot}$ and $\varepsilon_m$ reads as
    \begin{align}
    &\gamma_{ij}\frac{d}{dt}\Bigl[Na^3\Bigl(H^2+\frac{k}{a^2}
    -\frac{\varepsilon_m}{3M_P^2}\Bigr)\Bigr]\nonumber\\
    &\quad=N^2a^3H\Bigl[-G_{ij}+\Bigl(w\,G_{\bot\bot}
    -\frac{\dot\varepsilon_m}{3M_P^2}\Bigr)\gamma_{ij}
    \Bigr]=0,
    \end{align}
and vanishes in virtue of the matter stress tensor conservation law, $\dot\varepsilon_m=-3(\varepsilon_m+p_m)NH$, and Eq.(\ref{Eeij}). Therefore we get as an integral of equations of motion the GUMG Friedmann equation  (\ref{Eem}) with GUMG fluid density $\varepsilon=CM_P^2/Na^3$.

\bibliography{INFL_GUMG_revtex}

\end{document}